# Scaling up Reversible Logic with HKI Superconducting Inductors

Erik P. DeBenedictis
*Zettaflops LLC*
Albuquerque, New Mexico, USA
erikdebenedictis@gmail.com
April 11, 2025

*Abstract*—Researchers developed about a dozen semiconductor reversible (or adiabatic) logic chips since the early 1990s, validating circuit designs and proving the concept—but scale up required a further advance. This document shows that cryogenic inductors made of a new High Kinetic Inductance (HKI) material provide the advance. This material can be deposited as an integrated circuit layer, where it has enough energy recycling capacity to power a reversible circuit of the same size. This allows a designer to replicate and scale a complete reversible logic subsystem in accordance with Moore's law.

*Keywords—reversible logic, adiabatic logic, reversible computing, 2LAL, energy recycling power supply, CMOS, power-clocks, resonator, inductor, high kinetic inductance (HKI), YBCO, superinductor, intellectual property (IP)*

## I. EXECUTIVE SUMMARY

Reversible logic circuits created with funding from DARPA[1] and others used energy recycling to raise the energy efficiency of logic above CMOS levels, yet extending these efficiency gains to the chip level requires a scalable energy recycling power supply for the circuits. With the recent availability of HKI inductors for the power supply, it is now possible to create much more energy efficient quantum computer control chips for operation at 4 K, essentially by replacing the gates in cryo CMOS circuits (e.g. Horse Ridge[2]) with reversible circuits of higher energy efficiency, thus enabling more qubits before exceeding cryocooler capacity. This opportunity also applies to Read Out Integrated Circuits (ROICs), or sensors with *in situ* data processing.

HKI inductors now available operate at 4 K and below, but the approach may extend to 77 K using high-Tc superconducting inductors, such as Yttrium barium copper oxide (YBCO).[3-4] Thus, this approach may apply to DARPA's Low Temperature Logic Technology (LTLT) program,[5] which is exploring energy efficiency improvements for a broad range of applications in a cryogenic data center.

MIT-LL[6] demonstrated a hybrid monolithic foundry process with semiconductors (CMOS) and a Josephson junction process including HKI, while SeeQC[7] developed the HKI process used as an example in this document.

As a next step, this document suggests a scale up test of self–contained reversible logic IP ("intellectual property," which is a term of art in the semiconductor field for chip layout geometry), including CMOS-converted-to-reversible semiconductor circuits as a base and a superconducting layer that includes HKI. This IP would not require an additional energy recycling power supply, but could be AC-powered like three–phase electrical power transmission (but two- or four-phases at a lower voltage and higher frequency).

Multiple units of such IP could be laid out next to each other on a chip, leading to scalability rules similar to integrated circuits—rather than the less attractive scalability rules that apply to 3D structures—as was the case with previous reversible logic projects.

## II. INTRODUCTION

This document uses the result of a recent materials science advance called High Kinetic Inductance (HKI) to make reversible logic scalable. DARPA funded reversible logic in the early 1990s,[1] leading to about a dozen projects creating "successful" test chips. While successful, the test chips did not lead to R&D for scale up testing. In retrospect, the problem was that there was no material available at the time that had the required speed, power density, and scalability for an essential component called the energy recycling power supply. With HKI, it is now possible to fabricate scalable hybrid semiconductor-superconductor chips containing both reversible logic and its unique power supply.

*The benefit of reversible logic.* CMOS operates on DC power, where energy enters a chip on a DC power wire and charges transistor gate capacitance and capacitance between signal wires. When a data value changes, the $\frac{1}{2}CV^2$ energy on the capacitance turns into heat within the chip. The CMOS approach therefore turns $CV^2$ energy into heat for every signal change—and signal changes generally correspond to logic operations and are proxies for computational throughput.

The reversible logic approach uses an AC power supply. On the first half of each clock cycle, approximately the same amount of energy ($\frac{1}{2}CV^2$) used by CMOS enters the chip, but a reversible logic chip returns almost all the energy on the second half of each cycle—thus decreasing wall-power energy consumption by factors in the range of 10x-1,000x.

*The HKI advance.* Most reversible logic R&D comprises reversible logic circuitry and a separate energy recycling power supply. A semiconductor foundry creates the logic circuit, which is scalable much like a CMOS chip, but the energy recycling power supply relies on a non-scalable component, such as an inductor elsewhere on a printed circuit board or in a 3D structure. In other cases, it is an integrated component, such as a spiral inductor, but which is much larger than the reversible circuitry it supports.

A single layer of HKI inductors has about the same energy density as a cryo semiconductor circuit (CMOS or reversible, which have about the same energy density), thus allowing the non-scalable component in the previous paragraph to become an additional layer right on top of a reversible circuit.

*Potential impact.* Reversible logic is like a "Moore's law booster," meaning it applies to the latest semiconductor process at any point in time, increasing energy efficiency over and above what Moore's law provides. As such, reversible logic can increase performance within existing power envelopes or reduce power for the given application. Subject to cryogenic operating temperatures, reversible logic applies to applications that would otherwise use the latest version of CMOS.

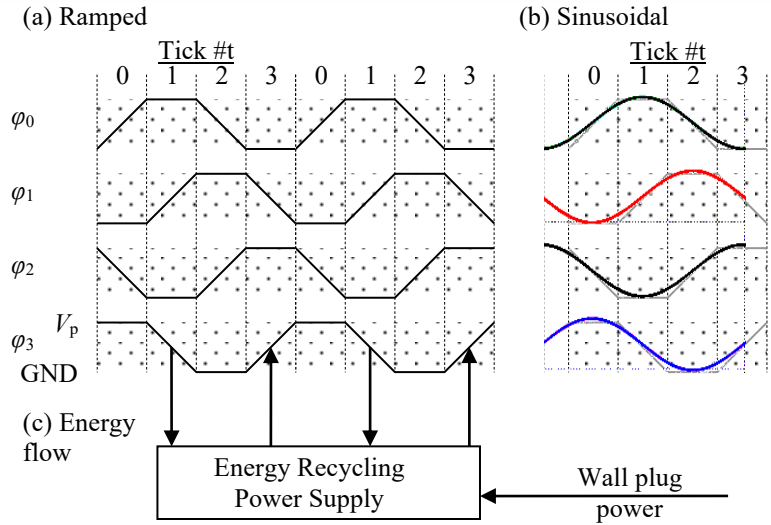

Fig. 1. (a) Ramped power-clocks, (b) sinusoidal power-clocks, (c) 4LC circuit, (d) module. Each phase is called a "tick."

With today's technology, reversible logic applies to 4 K cryogenic systems, such as quantum computer control and integrated sensor arrays. There is a future path for high Tc, 77 K operation, such as the DARPA LTLT program.[5]

*Potential scale-up test.* This document is not intended as an R&D proposal, but the technology appears ready for a scale-up test comprising the following phases:

1. Create and test (1a) basic test structures for a cryo semiconductor chip interacting with HKI inductors of approximately the correct inductance and current handling capacity and (1b) an initial version of a reversible logic circuit powered by an external supply.

2. Demonstrate (2a) a second version of the reversible semiconductor circuit, now of a design sufficient to draw conclusions about the scalability of reversible logic in general, powered by (2b) a second set of HKI inductors optimized for the specific circuit (2a).

III. TECHNICAL ISSUES IN SCALING UP REVERSIBLE LOGIC

This section explains how a High Kinetic Inductance (HKI) material enables reversible logic scale up. For brevity, this document does not describe the underlying principles of reversible logic, but interested readers can read ref. 8.

*Background.* Athas,[1] et. al. created, fabricated, and tested the AC-1 reversible microprocessor in the early 1990s under DARPA sponsorship, demonstrating increased energy efficiency over and above Moore's law. One might have expected follow-on DARPA scale up testing, but there was none. Instead, other R&D groups tested about a dozen variants of the circuits used in the DARPA project, coining circuit family names such as SQRL,[8] 2LAL,[9-10] RERL,[11] S2LAL,[12] Q2LAL,[13] and others—but the author does not know of any scale up tests. The author's position is that scale up required a physical science advance, so further circuit testing was destined to yield similar, unscalable results.

*Power-clock background.* Reversible logic uses energy recycling, as illustrated in Fig. 1. Past R&D on reversible circuits yielded many logic families, all with ramped 4- or 8-phase clocks. The 4-phase version illustrated in Fig. 1a is for 2LAL.[9-10] As illustrated in Fig. 1c for the bottom clock waveform, energy moves into the reversible circuit during the rising edge of the clock. The energy mostly charges gate, wire-to-GND, and wire-to-wire capacitance. The capacitive energy moves back into the energy recycling power supply during the falling edge of the clock, and the cycle repeats using the recycled energy plus a small amount of wall plug power.

The efficiency of the energy recycling process depends on resistive losses in the semiconductor circuit, which rise with clock frequency. Excluding the energy recycling power supply (because this document offers an improvement), the recycling efficiency can be around 90% at the typical clock rate of room-temperature CMOS (~1 GHz), indicating an energy efficiency increase of 10x. At lower speeds, the recycling efficiency may be 99% or higher, corresponding 100x or more.

The ramped clocks in Fig 1a are for the reader's convenience when accessing the large body of existing literature on reversible logic. In this document, the energy recycling power supply is just inductors, so the natural clocks are the sinusoidal waveforms in Fig. 1b. The reader will see the waveforms have the same basic shape – and the author's tests indicate they both have about the same energy efficiency when powering reversible circuits.

*Scale up story narrative.* To understand scale up, we first need a few key facts about the solution in Fig. 2c. The solution will be to cover the reversible logic semiconductor circuitry (bottom) with one or more layers of HKI material (top). Lithographic patterning will divide the HKI material into inductors (shown as meanders) that connect pairs of clock phases through the vertical pillars (the figure is an exploded view; the pillars are of near-zero height). An external non-recycling waveform will drive the clocks, but will draw an order of magnitude less energy from the wall plug than enters the circuit via the pillars. This is the energy efficiency advantage.

Fig. 2a illustrates a typical scenario for previous R&D efforts.[14] An R&D team created a reversible logic test chip as shown, in this case including a "power-clock generator" as identified by a blue arrow. Unfortunately, the R&D team was not funded for scale up testing even though the circuit testing was successful. The red arrow points to the problem, which is a tiny inductor symbol outside the chip. The chip may scale up according to Moore's law, but the inductor does not. Furthermore, the R&D team did not label the tiny inductor as "future HKI inductor" because HKI material was too new.

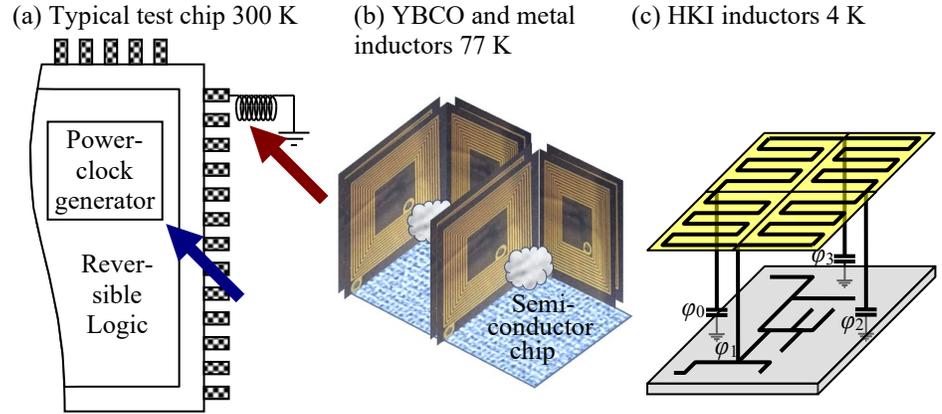

Fig. 2. (a) Typical reversible logic test chip. Includes a "power-clock generator" but also requires external inductors. (b) Geometric inductors requiring empty space. (c) HKI inductors. Exploded view; actual fabrication would be monolithic.

Fig. 2b illustrates a higher temperature but more challenging option based on non-HKI (also called non-superconducting or "geometric") inductors. These inductors require empty space to hold a magnetic field, such as the inside of a coil inductor or free space above and below an on-chip integrated spiral inductor (like the cloud in Fig. 2b). Since it is not possible to deposit "empty space" onto an integrated circuit, Fig. 2b is a 3D structure, which will not scale as well as predicted by Moore's law.

Fig. 2b shows actual manufactured and tested YBCO inductors. These inductors and their foundry process are close to what is needed for reversible logic, see refs. 3-4 for details. While the 3D structure is not an integrated circuit, there are options for high Tc inductors,[15] but they are beyond the scope of this document.

Fig. 2b also applies to normal metal (e.g. room temperature) electrical inductors. Aside from the 3D scalability issues just discussed, normal metal inductors are lossy at high frequencies. Further discussion of normal metal inductors is beyond the scope of this document.

We must now discuss how to implement Fig. 2c and show that an implementation can have sufficient energy capacity at speed.

*Converting circuits from CMOS to scalable reversible logic.* CMOS and reversible logic circuits are both comprise logic gates and wires, so a designer can replace CMOS logic circuits with the equivalent reversible logic circuits. This is not precisely true, but "true enough" for this argument. The reversible logic layout retains the long wires (e.g. busses) between groups of gates. Based on this imprecise argument, CMOS and reversible circuits will have about the same number of transistors and total wire length—hence the same amount of circuit capacitance and average energy flow during operation.

While the average energy flow rate is the same, the timing differs. Energy enters a CMOS circuit at a constant rate whereas Fig. 1c shows the energy entering the reversible circuit via power-clock $\varphi_n$ during tick $n$ (and leaving on tick $n+2$). The flow pattern for the sinusoidal clocks in Fig. 1b is conceptually the same but less distinct.

Numerically, the 4 mm x 4 mm Horse Ridge cryo CMOS qubit controller[2] consumes 10-140 mW, or $E_{CMOS}$ = 62-875 mW/cm$^2$. Based on the text, the range 10-140 mW corresponds to clock rates of $f$ = 100 MHz-1.6 GHz.

The energy capacity per unit area of an HKI material has a simple formula, specifically ½$L_\square I_{max}^2$, where $L_\square$ is the inductance per square of the process and $I_{max}$ is the maximum current per unit wire width. (To be clear, cutting a square centimeter of HKI material into wires of 0.1 mm and 0.01 mm width yields inductors with different inductances but nearly the same energy capacity.) For the SeeQC process,[7] $L_\square$ = 8.5 pH/$\square$, $I_c$ = 2.5 mA/μm = 2500 A/m, so ½$LI^2$ = 0.295 nW/cm$^2$. At $f$ = 1.6 GHz, $E_{reversible}$ = 472 mW/cm$^2$.

Based on rough numbers above, two layers of HKI inductor will suffice for powering a Horse Ridge-like chip converted to reversible logic.

State-of-the-art data center CPUs and accelerator chips have higher energy density and would require a hundred or more HKI layers. Theoretically, HKI inductors can operate in tall stacks like Flash memory, although tall HKI stacks are not currently manufacturable. HKI material is an active research area, so somebody might discover an HKI material with far higher energy density that would support high performance logic with fewer HKI layers.

*Fabricating the necessary structure.* Fig. 3 shows the baseline stack. The purple layers in Fig. 3a are an MIT-LL Josephson junction process stack (obtained from the internet) and the blue layers in Fig. 3b are the open source 130 nm Sky130 semiconductor process. After fabricating the semiconductor stack, the foundry planarizes the wafer and adds the superconductor stack using a Back End of Line (BEOL) process. Various foundries claim to have processes that are variants of Fig. 3, including MIT-LL.

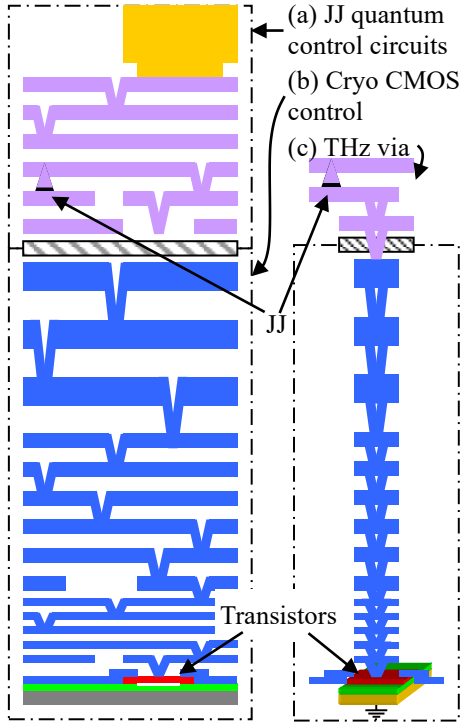

Fig. 3. JCMOS for quantum applications
(a) Back-End-Of-Line superconducting process on (b) standard CMOS process.
(c) Novel THz or SFQ-capable via.

This document does not use Josephson junctions, but just HKI layers and vias, the latter shown in Fig. 3c. It is possible that hybrid chips for scale up testing could use the stack in Fig. 3a, but where an organization interested in Josephson junctions funds the development of the fabrication process. Once the process is available, enterprising chip designers can use it for other use cases, including this one.

*Proof of design principle.* We now know that HKI inductors have sufficient energy capacity, but we need a "proof of principle" that they will work in a plausible circuit. Fig. 2a needed to minimize the number of external inductors because they limited scalability, but this new design point can have as many lithographically defined inductors as will fit in the available space.

Fig. 4a illustrates the circuit concept, coined 4LC (4-ell-cee), comprising four inductors that naturally oscillate as four sine waves in quadrature as shown in Fig. 1b. Each of the four dotted squares in Fig. 4b effectively contains the circuit in Fig. 4a with each $P_n = \varphi_n$ connected to the $C_n$ capacitive loading of the $P_n$ clock for the circuitry within the square. The array can be made as large as necessary, as long as the inductors around the boundary are applied in the pattern shown. The reader will note the absence of synchronization logic, high-current switches, and other complexities. The resonant modes of the LC array perform these functions naturally.

*Simulation.* The author created a simulation of the 4LC quad resonator in Fig. 4a, with the output appearing in Fig. 5.

For each plot in Fig. 5, the author entered the 4LC circuit into a circuit simulator, using initial conditions to create the sine waves in quadrature shown in Fig. 1b. The author then connected a 2LAL[9-10] reversible logic circuit to the sine waves, and found that the circuits would initially function properly. The circuit's function is to produce eight-phase clocks (as needed for several other reversible logic families), which are initially clean waveforms in the plots.

The 4LC circuit has oscillation modes, each characterized by a frequency and an amount of energy. Theoretically, the 4LC circuit is lossless and would oscillate forever. However, losses in the 2LAL circuit cause the total energy to decrease over time, including loss of energy in oscillation modes plus crosstalk between modes.

If the inductors and capacitors have low loss (i.e. the 4LC resonator has high Q), the circuit simulator will power the reversible logic for some time based on energy provided by the simulator's initial conditions. The energy will decline over

(a) Quad resonator called 4lc

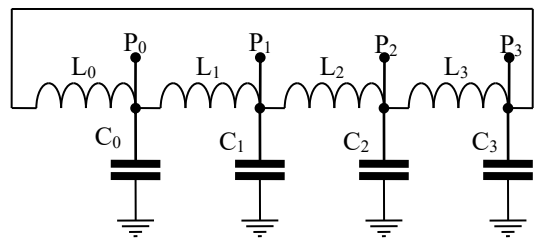

(b) Layout

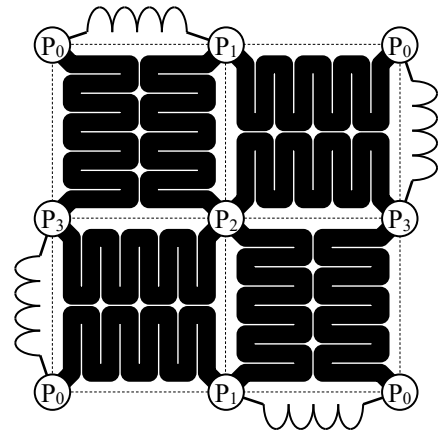

Fig. 4. (a) The 4lc quad resonator, which creates all four clocks at once with the correct phasing. (b) A potential design of the HKI layer, comprising a checkerboard of the quad resonators with clock phases labeled $P_0$-$P_3$. The array can extend to n x n and requires inductors double the value on the boundary as shown.

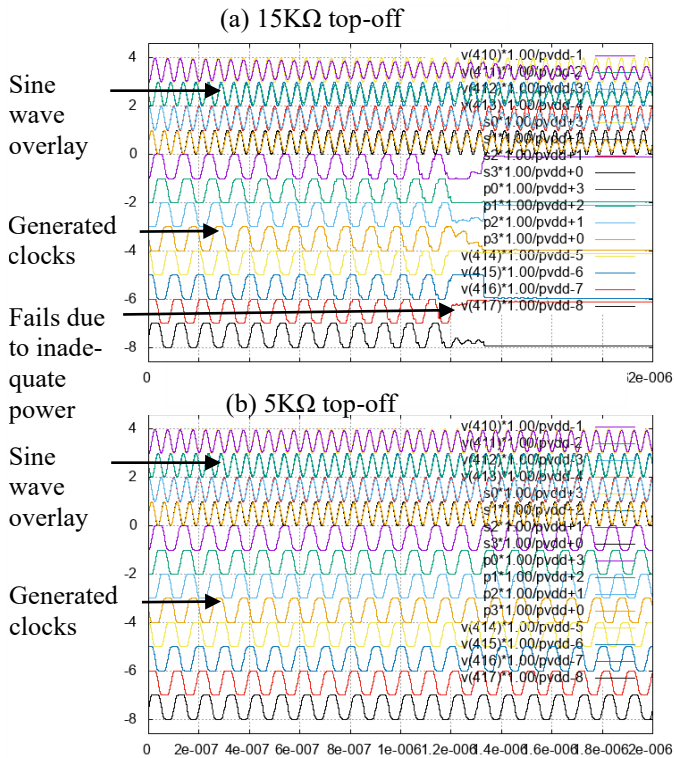

Fig. 5. Simulation of 4lc. (a) Works, runs down, and then fails. (b) "Topping off" allows continuous operation.

time, resulting in less amplitude in the sine waves and waveform distortion due to crosstalk. The "sine wave overlays" in Fig. 5a have two curves, the 4LC output values $P_n$ and a simulator-generated reference wave. The reader can see the $P_n$ amplitudes decline over time. As called out by the text "fails due to inadequate power" in Fig. 5a, the reversible logic circuit eventually produces a DC voltage. The remedy is to use external power to "top off" the energy in the desired oscillation mode while "draining out" undesired oscillation modes.

One way to "top off" the 4LC circuit would be to connect conventional sine waves, like those used in three-phase AC power, to the circuit through a resistor. The idea is that the resistor would transfer energy in or out of each oscillating mode until it matches the drive waveform's amplitude and phase. This includes topping off the desired mode and draining the energy from other modes. The first plot in Fig. 5a uses a 15 KΩ drive resistor, but the resistor was too weak and the circuit failed. However, the second plot in Fig. 5b uses a stronger 5 KΩ resistor, topping off the circuit adequately and yielding consistent sine waves and eight-phase clocks across the chart.

The author used two forms of AC power. The first form was four-phase AC power (shown in Fig. 1a). The second form was two-phase AC power, with phases 90° from each other. The two-phase version requires just three wires, GND and the two phases.

*Summary.* Reversible logic has a long history, but progress stalled due to a missing but necessary physical science advance. This document explains the advance and suggests the time is right for scale up testing.

The document showed how a cryo CMOS-translated-to-reversible-logic circuit and one or two layers of HKI inductor, both of the same area, have about the same energy content and jointly form an energy-recycling power supply. Since the resulting structure is entirely defined by chip layout geometry for a hybrid process, the result scales as predicted by Moore's law.

The document described and simulated a circuit called 4LC, which creates the four-phase energy recycling clocks used by 2LAL.[9-10] A non-energy recycling AC power waveform drives the 4LC.

With today's technology, reversible logic should be practical for 4 K cryogenic systems, such as quantum computer control and integrated sensor arrays. There is a future path for high Tc, 77 K operation, such as the LTLT program.

## IV. ADDITIONAL INFORMATION

The author has two additional files. If this publication (arXiv?) supports supplementary data, it will be added to this paper. The two files are:

A spreadsheet with equations for the top-level planning of reversible logic IP based on properties of the initial CMOS circuit and HKI parameters such as $L_\square$, $I_{max}$, and lithographic minimums for the width and space between HKI features. The spreadsheet explains how to scale the layout in Fig. 4b so the reversible logic operates at a specific clock rate—a topic not addressed in this document.

The ngspice simulation script that generated Fig. 5.

## REFERENCES

[1] Athas, William C. "Energy-recovery CMOS." *Low Power Design Methodologies*. Boston, MA: Springer US, 1996.

[2] Pellerano, Stefano, et al. "Cryogenic CMOS for Qubit Control and Readout." *2022 IEEE Custom Integrated Circuits Conference (CICC)*. IEEE, 2022. DOI: https://doi.org/10.1109/CICC53496.2022.9772841 https://pure.tudelft.nl/ws/portalfiles/portal/122719163/Cryogenic_CMOS_for_Qubit_Control_and_ReadoutTaverne.pdf.

[3] Brandl, Matthias F., et al. "Cryogenic resonator design for trapped ion experiments in Paul traps." *Applied Physics B* 122 (2016): 1-9. https://link.springer.com/content/pdf/10.1007/s00340-016-6430-z.pdf https://arxiv.org/abs/1601.06699.

[4] Brandl, Matthias F., et al. "Innsbruck high-temperature superconducting resonator." https://www.traphub.org/electronics/high_temperature_superconducting_resonator_ibk/ibk_hts_resonator.html.

[5] Low Temperature Logic Technology, web page https://www.darpa.mil/research/programs/low-temperature-logic-technology

[6] Superconducting integrated circuits, web page https://www.ll.mit.edu/research-and-development/advanced-technology/microsystems-prototyping-foundry/superconducting

[7] Yohannes, Daniel, et al. "Materials and methods for fabricating superconducting quantum integrated circuits." U.S. Patent No. 11,991,935. 21 May 2024. https://patents.google.com/patent/US11991935B2/en.

[8] Frank, Michael Patrick, and Thomas F. Knight Jr. *Reversibility for efficient computing*. Diss. Massachusetts Institute of Technology, Dept. of Electrical Engineering and Computer Science, 1999. https://dspace.mit.edu/handle/1721.1/9464.

[9] Athas, W. C., et al. "A framework for practical low-power digital CMOS systems using adiabatic-switching principles." *International Workshop on Low Power Design*. 1994.

[10] Anantharam, Venkiteswaran, et al. "Driving Fully-Adiabatic Logic Circuits Using Custom High-Q MEMS Resonators." *ESA/VLSI*. 2004. https://citeseerx.ist.psu.edu/document?repid=rep1&type=pdf&doi=7c8d654ce333d5b032af809c4c7770a61fb3add9

[11] Lim, Joonho, Dong-Gyu Kim, and Soo-Ik Chae. "Reversible energy recovery logic circuits and its 8-phase clocked power generator for ultra-low-power applications." *IEICE transactions on electronics* 82.4 (1999): 646-653. https://s-space.snu.ac.kr/bitstream/10371/21101/1/Reversible%20Energy%20Recovery%20logic%20circuits%20and%20its%208-phase%20clocked%20power%20generator%20for%20ultra-low-power%20applications.pdf.

[12] Frank, Michael P., et al. "Reversible computing with fast, fully static, fully adiabatic CMOS." https://doi.org/10.1109/ICRC2020.2020.00014 *2020 International Conference on Rebooting Computing (ICRC)*. IEEE, 2020. https://arxiv.org/pdf/2009.00448.

[13] DeBenedictis, Erik. "Managing energy in computation with reversible circuits." U.S. Patent Application No. 18/282,035. https://patents.google.com/patent/US20240152175A1/en.

[14] Kim, Suhwan, Conrad H. Ziesler, and Marios C. Papaefthymiou. "A true single-phase 8-bit adiabatic multiplier." *Proceedings of the 38th annual Design Automation Conference*. 2001.

[15] Srivastava, Yogesh Kumar, et al. "The elusive high-Tc superinductor." *arXiv preprint arXiv:2209.01342* (2022). https://arxiv.org/pdf/2209.01342.